# Emergence of bidirectional cell laning from collective contact guidance

Mathilde Lacroix[1], Bart Smeets[1,2], Carlès Blanch-Mercader[1], Samuel Bell[1], Caroline Giuglaris[1], Hsiang Yin Chen[1], Jacques Prost[1] and Pascal Silberzan[1*]

[1]Laboratoire PhysicoChimie Curie UMR168, Institut Curie, Paris Sciences et Lettres, Centre National de la Recherche Scientifique, Sorbonne Université, 75248 Paris, France
[2]BIOSYST-MeBioS, Katholieke Universiteit Leuven, Kasteelpark Arenberg 30, B-3001 Leuven (Heverlee), Belgium
*Correspondence to: pascal.silberzan@curie.fr

**Directed collective cell migration is central in morphogenesis, wound healing and cancer progression[1,2]. Although it is well-accepted that the molecular anisotropy of the micro-environment guides this migration[3,4], its impact on the pattern of the cell flows remains largely unexplored. Studying confluent human bronchial epithelial cells (HBECs) in vitro, we show that subcellular microgrooves elicit a polar mode of collective migration in millimeter-long bidirectional lanes that are much wider than a cell size, even though cell flows are highly disordered on featureless surfaces [5]. This directed flocking-like transition[6,7] can be accounted for by a hydrodynamic theory of active polar fluids and corresponding numerical simulations. This model further predicts that anisotropic friction resulting from the grooves lowers the threshold of the transition, which we confirm experimentally. Therefore, microscopic anisotropy of the environment not only directs the collective motion of the cells in the easy direction, but also shapes the cell migration pattern. Flow patterns induced by collective contact guidance are thus markedly different from those induced by supracellular confinement[8], demonstrating that all length-scales of the micro-environment must be considered in a comprehensive description of collective migration. Furthermore, artificial microtopographies designed from theoretical considerations can provide a rational strategy to direct cells to specific geometries and functions, which has broad implications, for instance, for tissue engineering strategies in organoid morphogenesis.**

Directed cell migration in development or disease is guided by the cell microenvironment such as neighbouring tissues, vessels or nerves[9,10] that delineate "highways" along which the cells can easily migrate. Cell motion can also be guided by aligned subcellular macromolecular bundles forming the extracellular matrix [11–14]. This anisotropy can be recapitulated in vitro with microfabricated substrates structured with micro-/nano-grooves[15]. In a wide range of groove geometries, isolated cells elongate, adapt their cytoskeleton organization, and migrate along the grooves in a process known as "contact guidance"[16,17]. In particular, because of its relevance to cancer cell migration[3,18], contact guidance of single cells has been thoroughly studied[11,19] in recent years.

The importance of directed collective migration at the scale of a cell population in processes such as wound healing or cancer invasion has been emphasized by many studies over the past decades[1,2,20,21]. This mode of migration goes beyond the superposition of individual behaviours and relies strongly on interactions between cells including excluded-volume interactions and cell-cell adhesion[22]. Surprisingly, the impact of oriented anisotropic subcellular cues on collective migration remains largely unexplored. It has, however, been shown that the healing of wounds in monolayers seeded on microgrooved substrates highly depends on the direction of the grooves relative to the migration front [23–26]. Such anisotropic substrates also strongly impact the collective migration of low activity epithelial and endothelial cells in confluent monolayers[27,28]. Here, we investigate the impact of collective contact guidance on a monolayer of cells that do not adhere together and whose activity is large enough to generate a chaotic velocity field on untextured substrates[5].

Our study therefore outlines the specificities of the various length-scales of the microenvironment on the collective migration of highly active cells. Furthermore, increasing our understanding of the impact of subcellular cues on collective migration can help design microstructures controlling spontaneous cell displacements and positioning. Applied to stem cells, this approach would open the way to new strategies for organoid morphogenesis

### Large scale laning of a cell monolayer on grooved substrates

To that end, we plated human bronchial epithelial cells (HBECs) on polydimethylsiloxane (PDMS) substrates that were either flat or microtextured with a subcellular castellated topographic pattern ("microgrooves") (Supplementary Figure 1). The direction of the grooves defines the $x$ direction (Supplementary Figure 1), and their period of 8 μm is smaller than a cell size (typically 20 – 30 μm in diameter). On a flat surface, HBEC monolayers developed active chaotic flows (Figures 1A,B), with a velocity correlation length of the order of 100 μm [5,30]. The distribution of the velocities then showed a preferred speed ($v = 38 \pm 12 \ \mu m/h$) with a random orientation (Figure 1C).

By contrast, when plated on a grooved substrate, the cells self-organized in wide lanes aligned with the grooves, in which they collectively migrated in antiparallel directions (Supplementary movie 1, Figures 1D,E). 20 h after confluence, the profile of the $x$ component of the velocity ($v_x$) along the $y$ direction alternated between $-v$ and $+v$ with $v = 45 \pm 14$ μm/h (n=7, N=4) (Figure 1F,H). Within a lane, all cells migrated at the same average velocity (Figures 1G,H; Supplementary Figure 2) with limited fluctuations in the $y$ direction. Between two adjacent lanes, the velocity changes sign over less than 50 μm (Supplementary Figure 2C), localizing the shear between two rows of cells.

We subsequently refer to this velocity pattern as the "laning" state. The lanes are millimeters long and hundreds of micrometers to millimeters wide (Figures 1G; Supplementary Figure 2A). To our knowledge, lanes of such large width have never been reported previously in boundary-free monolayers. Of note, we also observed an identical laning for cells seeded on adhesive/non-adhesive micropatterned lines with the same lateral dimensions as our grooves (Supplementary Figure 3), showing that the formation of the lanes is independent on the contact guidance mechanism.

Due to proliferation, cell density increased and the rms velocity in the monolayer decreased over time, independent of the structuration of the substrate (Supplementary Figure 4B). However, cells remained highly motile up to 25h post-confluence (Supplementary Figure 4). In the following, we focus on this motile regime ($t_{confluence} < t < t_{confluence} + 25h$). Of note, even though the monolayer started to jam 35h post-confluence [30,31], the spatial organization of the laning pattern did not change as long as velocities remained measurable (Supplementary Figure 4A): cells migrated more slowly but kept the same bidirectional velocity pattern.

To quantify the extent to which cells are guided by the grooves, we defined a guidance order parameter $S$ for the velocity orientation $S = \langle cos2\theta \rangle_{FOV}$, where $\theta$ is the angle of the velocity with respect to the grooves (Supplementary Figure 1D). $S = 0$ if velocity vectors are randomly oriented, whereas $S = 1$ when they are perfectly aligned with the grooves regardless of their direction. We found that on flat substrates, $S$ remained close to 0 during an experiment, whereas on grooved substrates, $S$ increased with time and eventually plateaued at $S = 0.89 \pm 0.01$ (Figure 2A). Therefore, collective cell guidance is enhanced as the laning pattern becomes more organized.

The lanes were characterized by the $v_x$ correlation length in the $x$ direction (see methods). On grooves, this correlation length increased with time, reaching several millimeters (Figure 2B) as the laning pattern became more organized (Supplementary Movie 1, Supplementary Figure 4A). Moreover, the average lane width also increased with time (Figure 2D). The laning pattern is therefore characterized by a time-increasing coordination in the parallel and perpendicular directions.

## Laning relies on cell polarity

Other types of alternating antiparallel streams guided by external cues have been reported in various active nematic systems ranging from microtubule bundles powered by kinesin motors [32] to endothelial cells (ECs) [27]. We argue, however, that the present situation is qualitatively different from these previous studies, and particularly from the streaming of ECs. Firstly, the width of EC streams is uniform, resulting in a well-marked periodicity of the velocity profile in the $y$-direction[27]. By contrast, the lane pattern observed with HBECs was aperiodic as demonstrated by the exponential distribution of the lane widths (Figures 1H, 2C). Secondly, the shapes of the $v_x$ profiles along the $y$-direction were very different, with ECs exhibiting sinusoidal velocity profiles[27] and HBECs displaying square profiles (Figure 1H, Supplementary Figure 2). Thirdly, nematic streaming in EC monolayers depends critically on cell-cell adhesion whereas HBECs are only weakly cohesive[30]. This last observation is consistent with the scattering of HBECs at the free edge of a monolayer (Supplementary Figure 5A) and with the cytoplasmic localization of the E-cadherins that ensure cell-cell adhesion of HBECs (Figures 3A, Supplementary Figure 5). Moreover, E-cadherin knockdown HBECs displayed the same laning pattern as the wild type cells (Figure 3B,C, Supplementary Figures 5B-E). Finally, whereas active nematic streaming of ECs is fully abolished by inhibition of contractility [32,33], inhibiting myosin II in the present experiments homogeneously reduced the value of $v_{rms}$, but did not significantly impair laning (Figure 3B,D; Supplementary Figure 6).

These results therefore indicate that HBEC laning differs fundamentally from nematic streaming and that HBEC monolayers can be described as active polar fluids for which activity is expressed through traction forces and not orientation gradients. To confirm this conclusion, we targeted the polarity of the cells with CK666, which inhibits the Arp2/3 complex that is an essential component of lamellipodial protrusions [34]. The addition of CK666 not only resulted in the expected decrease of the rms velocity (Figures 3E,F), but it also significantly decreased the $v_x$ correlation length in the $x$ direction (Figure 3H). However, the guidance order parameter retained a relatively high value ($S > 0.6$) compared to the flat situation (Figure 3G), suggesting that contact guidance is still effective when polarity is reduced. We therefore conclude that HBEC laning relies on protrusive activity and, hence, on cell polarity.

## A directed form of flocking transition

Based on these observations, we built a hydrodynamic continuum theory in which HBEC monolayers were described as an active polar fluid[35,36] characterized by its velocity and polarity fields (Supplementary Notes 1-4). The velocity results from the balance between internal stresses and cell-substrate force density that includes active traction forces as the dominant active process. Due to the presence of a cell-substrate interface, we included in the dynamics of the velocity field a velocity-polarity alignment term favouring the alignment of cell polarity with the velocity[37,38]. For the sake of simplicity, the grooves were modelled by an anisotropic friction only, although other terms of the equations could also be anisotropic. Specifically, we considered a larger friction coefficient along the "hard direction" (y-axis) compared with the "easy direction" (x-axis), in agreement with single cell behaviour (Supplementary Figure 1C).

To study how anisotropic friction influences the instability of a disordered state, we performed a linear stability analysis close to a disordered state (Supplementary Notes 2). In agreement with previous studies[38–40], on a homogeneous substrate, with isotropic friction, we found a disorder-to-order transition akin to a flocking transition controlled by an interplay between the velocity-polarity alignment, the active traction forces, and the friction coefficient (see Figure 4A when the friction anisotropy is 1). Introducing an $x - y$ friction anisotropy in the equation had three main consequences. First, above the disorder-to-order transition, a bidirectional laning mode was selected with flows along the easy direction (Figure 4A,C). Second, the mechanism of the instability occurred at longer wavelength, meaning that, as the laning pattern set in, the linear dynamics didn't select a characteristic wavelength. Thus, the emergent pattern was aperiodic (Figure 4B and Supplementary Notes 2). Third, increasing the friction coefficient along the hard $y$-direction reduced the effects of fluctuations and therefore lowered the laning instability threshold (see Figure 4A when the anisotropy is larger than 1 and Supplementary Notes 3).

This linearized continuum approach is valid only at the onset of instability close to the disordered state. To describe well-developed laning patterns, we conducted numerical simulations of self-propelled particles encompassing soft excluded volume interactions, anisotropic friction and velocity reinforcement $k_s$ which is the microscopic equivalent of the coarse-grained velocity-polarity alignment [41] (see Supplementary Notes 5). In the limit of isotropic friction, the velocity reinforcement promotes a flocking-like polar ordered state at high cell density [42]. These simulations predicted that anisotropic friction facilitates a transition from disorder to laning, which is in quantitative agreement with the continuum theory provided Gaussian fluctuations are included in the analysis (Figure 4D, Supplementary Movies 2&3, Supplementary Figure 8-10, Supplementary Notes 5.6). In well-formed lanes, particles were guided along the $x$-axis (low-friction, easy direction) (Figure 4D,E) and the $v_x$ profile along the $y$-direction alternated between two opposite values with no periodicity in the pattern of the lanes (Supplementary Figure 7A, 8). Finally, the exchange of cells in the hard direction appeared very limited compared to migration in wide tracks [43] (Supplementary Movie 2).

## Discussion

The continuum model and the numerical simulations based on the same ingredients therefore describe well the experimentally observed lanes and explain their emergence. They also provide testable predictions.

First, we note that, in these models, particles interact together only via excluded volume and not via cell-cell adhesion. This is consistent with our experimental observation that HBEC laning does not require cell-cell adhesion (Supplementary Figure 5). Furthermore, the theory hypothesizes a polar fluid or polar particles and doesn't explicitly include nematic terms. Indeed, we have shown that protrusive activity is central to the development of lanes (Figure 3E-H) whereas contractility is not (although it is essential in the migration process per se and controls the velocity within the lanes) (Figure 3D; Supplementary Figure 6).

In addition, the simulations conclude a non-periodic square velocity profile in the $y$-direction. More specifically, the distribution of lane widths is predicted to be exponential with a larger mean value on anisotropic substrates compared to isotropic substrates (Supplementary Figure 7A). The experiments indeed show a square profile and a similar exponential distribution of lane width (Figure 1H; 2C; Supplementary Figure 7B).

Theory and simulations show that friction anisotropy, $h$, is a control parameter of the transition (Figures 4A,D). We then posited that tuning this parameter could be recapitulated in the experiments by varying the groove depth $d$. $d = 0$ (flat surface) corresponds to an isotropic friction and hence to $h = 1$. In this case, the distribution of velocities is typical of a state with a preferred speed and random orientations. As the anisotropy increased, the velocity distribution became bimodal with two peaks located at opposite values on the $x$ axis (Supplementary Figure 8A). This evolution was fully recapitulated in the experiments as the groove depth $d$ increased (Supplementary Figure 9B). Additionally, we found that the square rms velocity $v_{rms}^2$ and the two normalized square components of the velocity $\frac{\langle v_x^2 \rangle}{v_{rms}^2}$ and $\frac{\langle v_y^2 \rangle}{v_{rms}^2}$ varied with groove depth in the experiments as they do with friction anisotropy in the simulations when the velocity reinforcement $k_s$ is sufficiently large (Supplementary Figure 10A,C ($k_s = 6$)). The continuum theory did not apply to this ordered regime but agreed with simulations for lower values of $k_s$ in the disordered regime (Supplementary Figure 10B,C ($k_s = 3$)). Along the same lines, the variations of the guidance order parameter $S$ with friction anisotropy in simulations, and as a function of groove depth in experiments were very similar (Figures 4F,G). To compare theory, simulations and experiments, we therefore chose to use $S$ as a common control parameter.

Both simulations and experiments showed a sharp transition of the average lane width for $0.2 < S < 0.4$ (Supplementary Figure 8D, Supplementary Figure 9D), confirming that substrate anisotropy lowers the disorder-to-laning transition threshold (Figures 4A,C). The slightly different shape of the transition in the simulations compared to the experiments in the ordered phase is attributed to the periodic boundary conditions implemented in the simulations (see also the differences between simulations and experiments in the shape of the lanes in the ordered regime (Supplementary Figure 8A, $h = 5$ vs. Supplementary Figure 9B, $d = 1.75 \mu m$)).

Simulations also predicted a near exponential increase of the $x$ correlation length as a function of the guidance order parameter (Figures 4H). Such a sharp variation was indeed observed in the experiments (Figure 4I) where the lanes' characteristic length reached a value of $1.9 \pm 0.3 \, mm$, for $S = 0.9 \pm 0.02$, which corresponds to a 10-fold increase compared to the value on a flat substrate ($200 \pm 60 \, \mu m$ for $S = 0$).

Modelling HBECs as an active polar fluid therefore accounts well for the breadth of experimental observations. However, previous experiments conducted on unpatterned substrates have concluded that HBEC monolayers could be described as a turbulent active nematic phase with half-integer topological defects[5]. To reconcile these seemingly contradictory findings, we analyzed how the local order scaled with the size $R$ of the window over which this order parameter was computed[44] (see Supplementary Notes 5.7). For randomly oriented $N$ particles, the order parameter scales as $\frac{1}{\sqrt{N}}$ as a consequence of the central limit theorem, and therefore as $\frac{1}{R}$ in two dimensions. Deviations from this scaling law indicate the development of some degree of order over macroscopic lengths.

We have quantified the orientational order from cell polarity on simulations of monolayers of polar cells in the disordered phase and in the absence of anisotropy (Supplementary Notes 5.7). For large values of $R$ in this disordered phase, all degrees of order were indeed described by the 1/R random scaling law. However, other degrees of order were also found at small R, with polar order predominating. In this regime, nematic order was the second most predominant order. In addition, the fluctuating polarity field and nematic field exhibited different typologies of defects characterized respectively by integer and half-integer topological charges (see Supplementary Notes 5.7). Therefore, cells that are individually polar may collectively organize with coexisting nematic and polar local orientational orders. This explains how polar HBECs plated on unpatterned substrates are described by a disordered nematic phase with half-integer defects[5], whereas they exhibit polar order on grooves.

## Conclusion

Altogether, we observe an excellent agreement between theory predictions backed by simulations, and experimental results. We thus conclude that the formation of lanes in HBEC cultures relies on three factors: cell self-propulsion, velocity-polarity alignment, and soft repulsion between neighbours. Therefore, the emergence of lanes in the cell monolayer above a critical groove depth is a form of flocking transition in a polar active fluid that is not only directed but also facilitated by collective contact guidance. Such mechanisms are generic and can be extended to other forms of anisotropies in the cells' environment; we therefore speculate that they might be at play in *in vivo* situations including tumour maturation[45] or morphogenesis[1].

Flow patterns generated by grooved surfaces are a good example of the emergence of flocking-like, large-scale, collective migration resulting from microscopic cues addressing each and every cell. These flows are distinct from the ones observed in supracellular wide tracks[8] or domains[46], where only the cells at the boundaries face a cue. Therefore, our study indicates that a global description of collective cell migration must encompass all length-scales of the cell environment. Furthermore, we propose that the pattern of the grooves can be rationally designed to position and orient the cells, thereby providing a way to engineer well-controlled situations prone to specific biological functions such as geometry-controlled stem cell differentiation or organoid morphogenesis[47,48].

## Acknowledgements

We thank the members of the Biology-inspired Physics at MesoScales (BiPMS) and the Physical Approach of Biological Problems (PABP) groups, and in particular Xinming Wu for his help with the SEM. The BiPMS group is an associate member of the Institut Pierre-Gilles de Gennes and has benefited from the technical contribution of the joint service unit CNRS UAR 3750. The Biologie Moléculaire et Biologie Cellulaire platform of the UMR168 is gratefully acknowledged and in particular Aude Battistella and Fanny Cayrac for their help with the transfection of the cells, and John Manzi for the western blots. We thank the Cell and Tissue Imaging core facility (PICT IBiSA), Institut Curie, member of the French National Research Infrastructure France-BioImaging (ANR10-INBS-04).

The BiPMS group and the PABP group are members of the LabEx Cell(n)Scale (grants ANR-11-LABX-0038, ANR-10-IDEX-0001-02). The authors acknowledge the support of the Agence Nationale de la Recherche (ANR), under grant ANR-18-CE30-0005.

HYC gratefully acknowledges funding from the Ministry of Science and Technology, Taiwan, under contract No. 109-2917-I-564-013, and from LabEx Cell(n)Scale. CG gratefully acknowledges funding from the Fondation pour la Recherche Médicale. BS acknowledges support from the Research Foundation Flanders (FWO), grant 12Z6118N, and KU Leuven internal funding C14/18/055. SB and JP acknowledge funding by the Human Frontiers in Science Program (HFSP RGP0038/2018)

## Author contributions

PS designed the research. ML, CG and HYC performed the experiments. ML, BS, CBM CG, HYC and PS analysed the experimental data. CBM, SB, BS and JP made the theoretical model. BS performed the numerical simulations and analysed the simulation data. All authors contributed to writing the manuscript.

## Competing interests

The authors declare no competing interests.

## References

1. Scarpa, E. & Mayor, R. Collective cell migration in development. *J. Cell Biol.* **212**, 143–155 (2016).
2. Friedl, P. & Gilmour, D. Collective cell migration in morphogenesis, regeneration and cancer. *Nat. Rev. Mol. Cell Biol.* **10**, 445–57 (2009).
3. Clark, A. G. & Vignjevic, D. M. Modes of cancer cell invasion and the role of the microenvironment. *Curr. Opin. Cell Biol.* **36**, 13–22 (2015).
4. Paul, C. D., Mistriotis, P. & Konstantopoulos, K. Cancer cell motility: lessons from migration in confined spaces. *Nat. Rev. Cancer* **17**, 131–140 (2017).
5. Blanch-Mercader, C. *et al.* Turbulent Dynamics of Epithelial Cell Cultures. *Phys. Rev. Lett.* **120**, 208101 (2018).
6. Vicsek, T., Czirók, A., Ben-Jacob, E., Cohen, I. & Shochet, O. Novel type of phase transition in a system of self-driven particles. *Phys. Rev. Lett.* **75**, 1226 (1995).
7. Giavazzi, F. *et al.* Flocking transitions in confluent tissues. *Soft Matter* **14**, 3471–3477 (2018).
8. Duclos, G. *et al.* Spontaneous shear flow in confined cellular nematics. *Nat. Phys.* **14**, 728–732 (2018).
9. Alexander, S., Weigelin, B., Winkler, F. & Friedl, P. Preclinical intravital microscopy of the tumour-stroma interface: invasion, metastasis, and therapy response. *Curr. Opin. Cell Biol.* **25**, 659–71 (2013).
10. Sarna, M., Wybieralska, E., Miekus, K., Drukala, J. & Madeja, Z. Topographical control of prostate cancer cell migration. *Mol. Med. Rep.* **2**, 865–871 (2009).
11. Thrivikraman, G. *et al.* Cell contact guidance via sensing anisotropy of network mechanical resistance. *Proc. Natl. Acad. Sci. U. S. A.* **118**, 1–11 (2021).
12. Chanrion, M. *et al.* Concomitant Notch activation and p53 deletion trigger epithelial-to-

mesenchymal transition and metastasis in mouse gut. *Nat. Commun.* **5**, 5005 (2014).
13. Erdogan, B. *et al.* Cancer-associated fibroblasts promote directional cancer cell migration by aligning fibronectin. *J. Cell Biol.* **216**, 3799–3816 (2017).
14. Sherwood, D. R. Basement membrane remodeling guides cell migration and cell morphogenesis during development. *Curr. Opin. Cell Biol.* **72**, 19–27 (2021).
15. Curtis, A. & Wilkinson, C. Topographical control of cells. *Biomaterials* **18**, 1573–1583 (1997).
16. Leclech, C. & Villard, C. Cellular and Subcellular Contact Guidance on Microfabricated Substrates. *Front. Bioeng. Biotechnol.* **8**, 551505 (2020).
17. Nguyen, A. T., Sathe, S. R. & Yim, E. K. F. From nano to micro: topographical scale and its impact on cell adhesion, morphology and contact guidance. *J. Phys. Condens. Matter* **28**, 183001 (2016).
18. Provenzano, P. P. *et al.* Collagen density promotes mammary tumor initiation and progression. *BMC Med.* **6**, 1–15 (2008).
19. Leclech, C. & Barakat, A. I. Is there a universal mechanism of cell alignment in response to substrate topography? *Cytoskeleton* **78**, 284–292 (2021).
20. Liu, R. *et al.* Diversity of collective migration patterns of invasive breast cancer cells emerging during microtrack invasion. *Phys. Rev. E* **99**, 062403 (2019).
21. Alexander, S., Koehl, G. E., Hirschberg, M., Geissler, E. K. & Friedl, P. Dynamic imaging of cancer growth and invasion: a modified skin-fold chamber model. *Histochem. Cell Biol.* **130**, 1147–54 (2008).
22. Hakim, V. & Silberzan, P. Collective cell migration: a physics perspective. *Reports Prog. Phys.* **80**, 076601 (2017).
23. Lee, G. *et al.* Contact guidance and collective migration in the advancing epithelial monolayer. *Connect. Tissue Res.* **59**, 309–315 (2018).
24. Dalton, B. a *et al.* Modulation of epithelial tissue and cell migration by microgrooves. *J. Biomed. Mater. Res.* **56**, 195–207 (2001).
25. Lawrence, B. D., Pan, Z. & Rosenblatt, M. I. Silk Film Topography Directs Collective Epithelial Cell Migration. *PLoS One* **7**, e50190 (2012).
26. Londono, C. *et al.* Nonautonomous contact guidance signaling during collective cell migration. *Proc. Natl. Acad. Sci.* **111**, 1807–1812 (2014).
27. Leclech, C. *et al.* Topography-induced large-scale antiparallel collective migration in vascular endothelium. *Nat. Commun.* **13**, 2797 (2022).
28. Haga, H., Irahara, C., Kobayashi, R., Nakagaki, T. & Kawabata, K. Collective movement of epithelial cells on a collagen gel substrate. *Biophys. J.* **88**, 2250–2256 (2005).
29. Kim, J., Koo, B. K. & Knoblich, J. A. Human organoids: model systems for human biology and medicine. *Nat. Rev. Mol. Cell Biol.* **21**, 571–584 (2020).
30. Garcia, S. *et al.* Physics of active jamming during collective cellular motion in a monolayer. *Proc. Natl. Acad. Sci.* **112**, 15314–15319 (2015).
31. Park, J.-A. *et al.* Unjamming and cell shape in the asthmatic airway epithelium. *Nat. Mater.* **14**, 1040–1048 (2015).
32. Guillamat, P., Ignés-Mullol, J. & Sagués, F. Control of active liquid crystals with a magnetic field. *Proc. Natl. Acad. Sci.* **113**, 5498–5502 (2016).
33. Thijssen, K., Metselaar, L., Yeomans, J. M. & Doostmohammadi, A. Active nematics with anisotropic friction: The decisive role of the flow aligning parameter. *Soft Matter* **16**, 2065–2074 (2020).

34. Hetrick, B., Han, M. S., Helgeson, L. A. & Nolen, B. J. Small molecules CK-666 and CK-869 inhibit actin-related protein 2/3 complex by blocking an activating conformational change. *Chem. Biol.* **20**, 701–712 (2013).
35. Lee, P. & Wolgemuth, C. Advent of complex flows in epithelial tissues. *Phys. Rev. E* **83**, 061920 (2011).
36. Blanch-Mercader, C. *et al.* Effective viscosity and dynamics of spreading epithelia: a solvable model. *Soft Matter* **13**, 1235–1243 (2017).
37. Brotto, T., Caussin, J. B., Lauga, E. & Bartolo, D. Hydrodynamics of confined active fluids. *Phys. Rev. Lett.* **110**, (2013).
38. Kumar, N., Soni, H., Ramaswamy, S. & Sood, A. K. Flocking at a distance in active granular matter. *Nat. Commun.* **5**, (2014).
39. Maitra, A., Srivastava, P., Marchetti, M. C., Ramaswamy, S. & Lenz, M. Swimmer Suspensions on Substrates: Anomalous Stability and Long-Range Order. *Phys. Rev. Lett.* **124**, 28002 (2020).
40. Heinrich, M. A. *et al.* Size-dependent patterns of cell proliferation and migration in freely-expanding epithelia. *Elife* **9**, 1–21 (2020).
41. Smeets, B. *et al.* Emergent structures and dynamics of cell colonies by contact inhibition of locomotion. *Proc. Natl. Acad. Sci.* **113**, 14621–14626 (2016).
42. Zhao, Y., Ihle, T., Han, Z., Huepe, C. & Romanczuk, P. Phases and homogeneous ordered states in alignment-based self-propelled particle models. *Phys. Rev. E* **104**, 044605 (2021).
43. Vedula, S. R. K. *et al.* Emerging modes of collective cell migration induced by geometrical constraints. *Proc. Natl. Acad. Sci. U. S. A.* **109**, 12974–9 (2012).
44. Eckert, J., Giomi, L. & Schmidt, T. Hexanematic crossover in epithelial monolayers depends on cell adhesion and cell density. *arXiv* (2022).
45. Friedl, P., Locker, J., Sahai, E. & Segall, J. E. Classifying collective cancer cell invasion. *Nat. Cell Biol.* **14**, 777–783 (2012).
46. Deforet, M., Hakim, V., Yevick, H. G., Duclos, G. & Silberzan, P. Emergence of collective modes and tri-dimensional structures from epithelial confinement. *Nat. Commun.* **5**, 3747 (2014).
47. Gjorevski, N. *et al.* Tissue geometry drives deterministic organoid patterning. *Science (80-. ).* **375**, eaaw9021 (2022).
48. Karzbrun, E. *et al.* Human neural tube morphogenesis in vitro by geometric constraints. *Nature* **599**, 268–272 (2021).

Figure 1

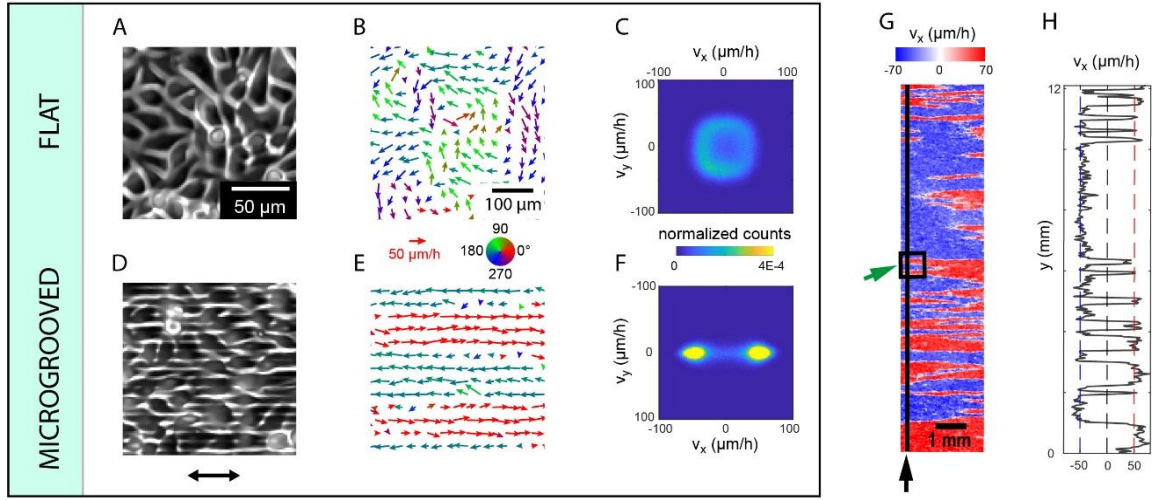

**Figure 1: Flows within HBEC monolayers on grooved substrates self-organize in a bidirectional laning pattern**

A, D: Phase contrast image of an HBEC monolayer on a flat (A) vs grooved (D) substrate. In D, the double-headed arrow shows the groove orientation.

B, E: Representative velocity fields on a flat (B) and grooved (E) substrate. The colour codes for the velocity orientation as shown. Note the laning pattern on grooved substrates (E) whereas flows are highly disordered on flat substrates (B).

C, F: Bivariate histograms of the two components of the velocity $(v_x, v_y)$ on a flat (C) and grooved (F) substrate. On flat substrates, the velocity has a preferred module while its direction is uniformly distributed. On grooved substrates, cell displacements are constrained along the direction of the grooves. Flat substrates, n=10; grooved substrates, n=7.

G, H: Representative large-scale cartography and profile of $v_x$ on a grooved substrate. The box (green arrow) corresponds to the velocity field shown in (E). The line (black arrow) corresponds to the velocity profile shown in panel (H). The x component of the velocity varies as a square, aperiodic signal.

All panels: t = 20h post-confluence.

Panels D-H: groove depth= 1.75 µm

Figure 2

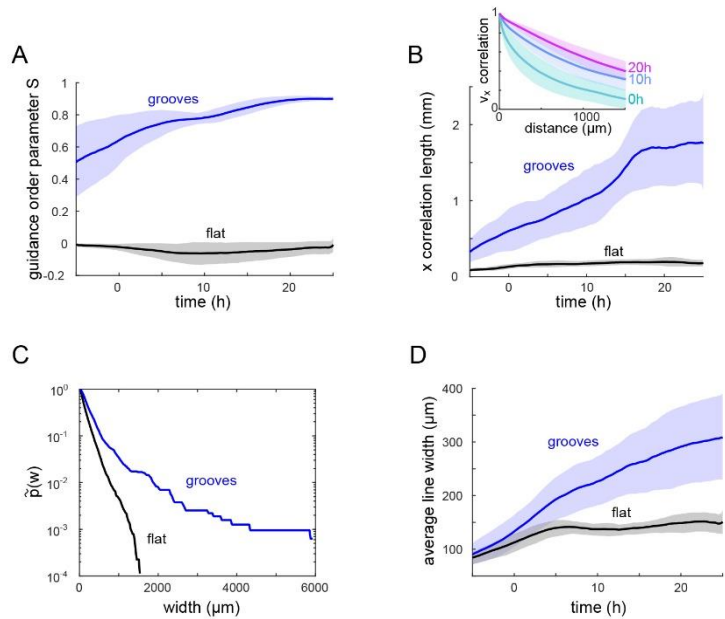

**Figure 2: Development of the laning pattern**

A: The guidance order parameter, $S$, of the velocity increases with time on grooved substrates until reaching a value close to 1 (blue line). No order is observed on flat substrates (black).

B: The correlation length of $v_x$ along $x$ increases with time on grooved substrates and can reach up to millimeter values (blue line), whereas it remains low on flat substrates (black line). Inset: $v_x$ correlation functions along $x$ on grooved substrates for 3 different times after confluence.

C: Cumulative distribution of lane width $\tilde{p}(w)$ on grooved (blue) vs flat (black) substrates (t=20h post confluence). The exponential distribution of the lane widths is a signature of the non-periodicity of the lanes pattern.

D: Coarsening of the lanes with time on grooved substrates (blue). The final mean width reaches several 100 µm before the cells jam.

All panels: groove depth = 1.75µm. Typical FOVs: 10 mm X 30 mm. t= 0 corresponds to confluence

Shaded areas correspond to the standard deviation over n= 10 (respectively n=7) FOVs for flat (respectively grooved) substrates.

Figure 3

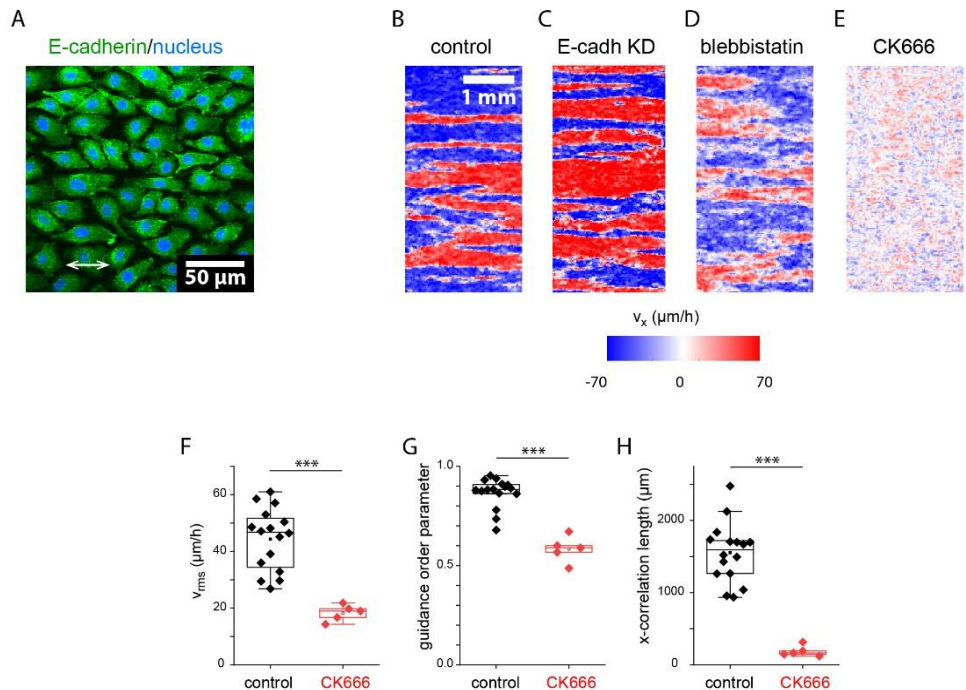

**Figure 3: The laning pattern relies on cell polarity but not on cell-cell adhesion nor on cell contractility.**

A: E-cadherin immunostaining of an HBEC confluent monolayer migrating collectively on a grooved substrate in a laning pattern. E-cadherins are mostly cytoplasmic, resulting in weak cell-cell adhesion.

B-E: $v_x$ laning pattern of HBECs on grooved substrates in different conditions.

B: Control experiment.

C: The well-established pattern obtained with an E-cadherin knocked down HBEC cell line confirms that cell-cell adhesion is dispensable in the emergence of laning.

D: Interfering with cell contractility by inhibiting myosin II with blebbistatin slows down the cells but doesn't prevent the emergence of the laning pattern.

E: Interfering with cell polarity with CK666, which inhibits Arp2/3 implicated in lamellipodium formation, prevents the emergence of a laning pattern.

F, G, H: impact of CK666 on HBEC laning. The average velocity $v_{rms}$ (F), the guidance order parameter $S$ (G) and the x correlation length (H) quantitatively decrease in presence of CK666. n= 16 (resp. 5) FOVs analyzed for control (resp. CK666) conditions. Boxplots (box = median and 25-75% confidence interval, whiskers = non-outlier range). *** denotes p<0.001 (two-sample t-test.)

All panels: t=20h after confluence; groove depth = 1.45µm

Figure 4

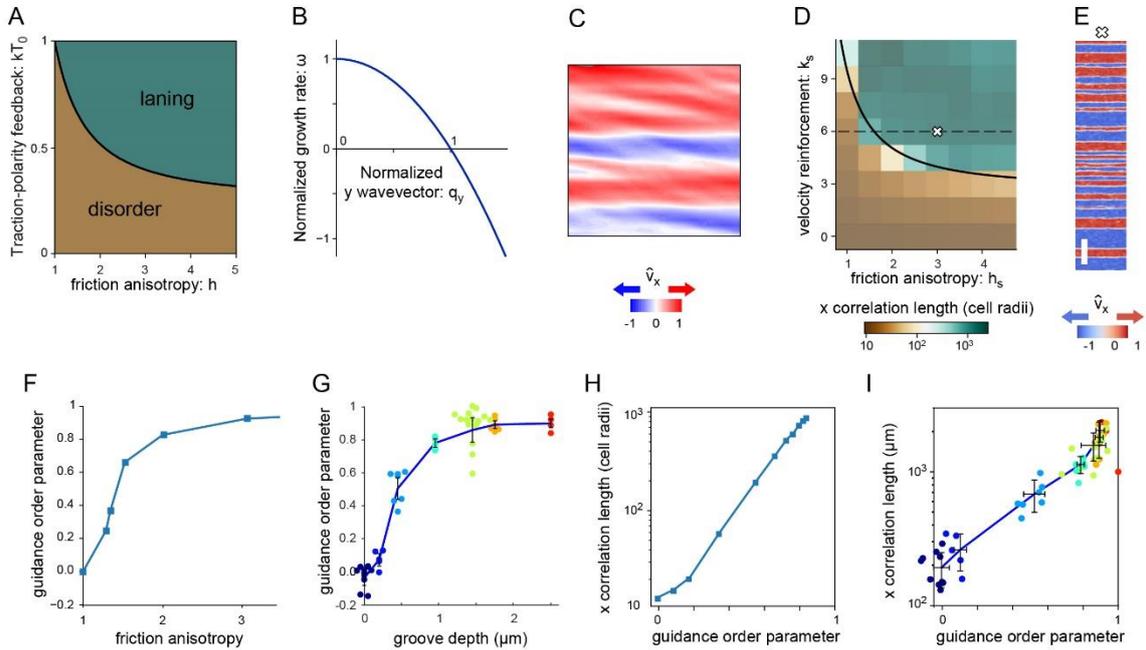

**Figure 4: Bidirectional laning of an active polar fluid results from anisotropic friction**

A-C: continuum theory of the disorder-to-laning transition in active polar fluids with anisotropic friction and traction-polarity feedback.

A: Transition (black line) between the disorder phase (brown) and the laning phase (green) as a function of the friction anisotropy and the traction-polarity feedback. The critical traction-polarity feedback decreases as the friction anisotropy increases.

B: Dominant growth rate for wavevector $q_x = 0$ as a function of $q_y$. The fastest mode of growth of the laning pattern is obtained for $q = 0$ and therefore the emergent pattern is aperiodic. The growth rate and wavevector $q_y$ are normalized and plotted above threshold - shear viscosity $= 0$.

C: Velocity field at the onset of the laning instability. Colours code for the x-component of the velocity. Integration domain is set to L=800. Panel size is 400.

D – E: Numerical simulations of self-propelled particles with anisotropic friction. Simulations are performed when reaching macroscopic timescales associated with coarsening.

D: Phase diagram of the correlation length $L_x$ of $v_x$ in the $x$ direction in the phase space (velocity reinforcement $k_s$, friction anisotropy $h_s$). High values of $L_x$ (green) correspond to laning (Supplementary Movie 3). Friction anisotropy favours laning by decreasing the critical value of $k_s$. The solid black line is the transition curve predicted by the continuum theory ($k_s \sim h_s^{-3/2}$) (see panel (A)).

E: Friction anisotropy guides the particles along the x axis and generates a bidirectional laning organization over large fields of view. This particular pattern was acquired at the white cross shown on panel D ($h_s$=3, $k_s$=6). See Figure 1G for comparison with experiments. Bar=50 cell diameters.

F, G: The increase of the guidance order parameter S with friction anisotropy $h_s$ in the simulations (F) is faithfully reproduced by varying the groove depth $d$ in the experiments (G). The colours of the points in (G) reflect the binned groove depth. Each measurement in (F) is the average of 5 repeats of the simulation.

H, I: The x-correlation length increases exponentially with the guidance order parameter in the simulations (H) as in the experiments (I). Each measurement is the average of 5 repeats of the simulation (H). The colour code in (I) refers to panel (G).

Experimental error bars are standard deviations (G, I). Lines are guides for the eye (F-I).

Panels G, I: t=20h post-confluence. See Methods for statistical details.

Panel F,H : Parameters of the simulations are listed in Supplementary Table 1. $k_s$ = 6.

## Methods

**Cell culture**

Human bronchial epithelial cells (HBECs) (gift from J. Minna's laboratory in Dallas, TX) were cultured in supplemented keratinocyte serum-free medium with L-glutamine (Keratinocyte-SFM with L-glutamine; Gibco).
Cells were maintained at 37°C under 5% $CO_2$ partial pressure and 95% relative humidity.

**Microfabrication and cell seeding**

Microstructured masters were fabricated from a photoresist layer spin-coated on silicon wafers using classical photolithography protocols[48] either by direct laser writing (µPG 101; Heidelberg instruments) or by UV illumination (MJB3 Mask aligner; Karl Süss). The photoresist (SU-8; MicroChem) was diluted to tune the final thickness. Grooved substrates of various depths were then obtained after UV exposure through a chromium mask and development. Wafers were then silanized in a vapour of (tridecafluoro-1,1,2,2-tetrahydrooctyl)trichlorosilane (abcr) for 2h and used as moulds to make polydimethylsiloxane (PDMS Sylgard 184, Dow Corning) replica. After being peeled off from the wafer, their surface was activated with a 30s air plasma. The PDMS slabs were then placed in a 12-well glass bottom plate (CellVis) and coated with a 10µg/mL fibronectin solution (Fibronectin Bovine Protein, Bovine Plasma, Gibco) for 1h. Unless otherwise stated, cells were seeded at high density (225 000 cells/$cm^2$ i.e. 900 000 cells per well) and were left to adhere to the substrate for a few hours (typically 6h). Prior to imaging, cell monolayers were rinsed with PBS and 5mL of fresh medium was added. For all experiments, the origin of time is taken when cells reach confluence (~15 h after seeding).
Alternatively, cells were plated on chemically micro-patterned substrates made of alternating 4 µm wide lines of poly(ethylene glycol) (PEG) and fibronectin following previously described protocols[48]. Briefly, the glass coverslip was uniformly coated with a covalently attached PEG layer. A layer of photoresist was spin-coated on top of the PEG coating and patterned by photolithography to define a photoresist mask. At that point, the PEG layer was selectively etched in the unprotected areas in an air plasma. After dissolution of the remaining resist in acetone and thorough rinsing with water, samples were incubated in a fibronectin solution.
For the wound healing assay[49] (Supplementary Figure 5), we used commercially available silicone-based Culture-Inserts 2 Well (Ibidi). Cells were seeded at high density in both chambers and left to incubate on fibronectin-coated glass coverslips for a few hours until they were fully attached and confluent at which point, the insert was removed. Cells were rinsed with PBS and fresh medium was added prior to imaging.

**Drugs and inhibitors**

Drugs were added approximately 6h after cell seeding in the fresh medium: CK666 (Sigma-Aldrich, SML0006) was used at 100 µM and para nitro-blebbistatin (Cayman Chemical) at 5 µM and 10 µM (concentrations larger than 10µM were toxic for the cells and therefore not used). Para nitro-blebbistatin is less photosensitive than blebbistatin and was therefore preferred, although for simplicity, herein we refer to it as "blebbistatin".

**E-cadherin knock down (KD) and western blot**

E-cadherin was knocked down through E-cadherin shRNA (h) lentiviral particles (Santa Cruz Biotechnology). Briefly, HBECs were incubated overnight in their medium supplemented with 5µg/mL polybrene (Santa Cruz Biotechnology) and E-cadherin shRNA (h) lentiviral particles. Cells expressing the shRNA were grown before being selected and further cultured with 10μg/mL puromycin dihydrochloride (Gibco). Knocked down efficiency was confirmed by Western blot analysis that showed a reduction of expression of 75-80%. (Supplementary Figure 11)

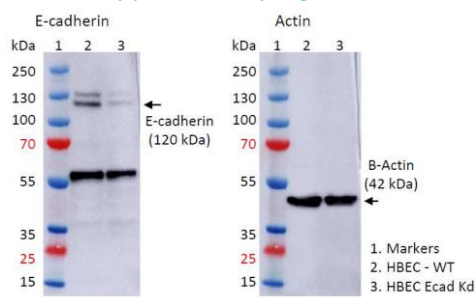

***Supplementary Figure 11***: *Western blots for E-cadherin (left) and actin (right) for WT HBECs (track 2) and E-cadherin KD cells (track 3). The expression of E-cadherin was successfully reduced in E-cadherin KD cells although the actin expression was unaffected.*

**Time-lapse microscopy**

Time-lapse multi-field experiments were performed in phase contrast on an automated inverted microscope (Olympus IX71) equipped with temperature, humidity and $CO_2$ regulations (Life Imaging service). With a 4X objective, the field of view (FOV) was 3.3 mm × 3.3 mm. The displacements of the sample and the acquisitions of images (camera Prime BSI, Teledyne Photometrics) were controlled by Metamorph (Universal Imaging) software. The delay between two successive images of the same field was set to 5 minutes. Larger fields of view were obtained by vertically stitching four FOVs in the center of the PDMS substrate. Acquisitions typically ran for 2 to 3 days. For the wound healing assays, time-lapse multi-field experiments were performed in phase contrast with a 10X objective.

**Fluorescence imaging**

Cells were first fixed with 4% (wt/vol) paraformaldehyde (Electron Microscopy Sciences), permeabilized with Triton X-100 (Euromedex), and saturated with 0.1% Bovine Serum Albumin (BSA; (Sigma)) and 2.5% Normal Goat Serum (NGS; ThermoFisher) in PBS. E-cadherin labelling was performed with a Rat anti-E-cadherin antibody (Invitrogen) coupled to a secondary Goat anti-Rat Alexa Fluor 488 antibody (Invitrogen). Nuclei were labelled with Hoechst 33342 (Molecular Probes)
Fixed and stained samples were imaged by fluorescence on an inverted spinning disk confocal microscope (CSU-W1, Roper/Nikon) using a 20X oil objective.

**Image analysis and data analysis**

Most of the image processing was performed using Fiji[50] public domain software[51]. Stitching of the FOVs was performed using the Pairwise Stitching plugin[52]. Stitched images were then rotated so that the grooves aligned with the horizontal direction, and cropped. Dimensions of the final images were typically 10 mm x 3 mm.
The velocity field was computed by particle image velocimetry (PIV) analysis[53]. Stacks of images were analyzed with a custom-made PIV algorithm based on the MatPIV software package[54] for Matlab (MathWorks). The window size was set to 16 pixels = 52 μm with a 0.5 overlap. Outlier vectors were removed with a global filter and replaced by data interpolated from neighbouring values. A sliding time average over 30 minutes (6 time frames) was then performed.

The guidance order parameter was defined as $S = \langle cos2\theta \rangle_{FOV}$, where $\theta$ is the angle of the velocity relative to the groove orientation.

The correlation function of the x component of the velocity, $v_x$, along the $x$ direction was defined as:

$$C_x(t, \boldsymbol{\delta x}) = \frac{\langle v_x(t, \boldsymbol{r}) \cdot v_x(t, \boldsymbol{r} + \boldsymbol{\delta x}) \rangle_r}{\langle v_x(t, \boldsymbol{r})^2 \rangle_r}$$

Where $\boldsymbol{\delta x} = \delta x \cdot \hat{\boldsymbol{x}}$, $\hat{\boldsymbol{x}}$ being the unit vector in the x direction.

The corresponding correlation lengths were measured from an exponential fit of these correlation functions.

The distribution of lane widths is accounted for by the probability p̃(w) that a lane is larger than w. The average lane width was averaged over 11 profiles for each FOV. The width of a lane is defined as the distance between the two points where the $v_x$ profile along the y direction crosses the $v_x$ = 0 axis.

Data analysis was performed with Matlab (Mathworks) or with Origin (Originlab).

**Statistics**

Each experiment was conducted with a 12-well plate, each well amounting to an independent experiment. Experiments were all physically replicated at least twice with at least two independent wells for each replica. Results were pooled from these different experiments. The error bars are the standard deviations. We note n the total number of analyzed wells over N replicas. These numbers are given in Supplementary Table 2 for experiments dealing with wild type cells in the absence of drugs, and are reported in the corresponding Figure captions:

| Groove depth $d$ (nominal value) (µm) | n | N |
|---|---|---|
| 0 (flat substrate) | 10 | 6 |
| 0.2 | 5 | 3 |
| 0.45 | 7 | 4 |
| 0.95 | 6 | 3 |
| 1.45 | 15 | 8 |
| 1.75 | 7 | 4 |
| 2.5 | 4 | 4 |

***Supplementary Table 2***: *Total number, n, of experiments (ie number of analyzed wells). N is the number of physical replicas.*

**Numerical method to solve the linear dynamics of an active polar fluid:**

In Supplementary Notes 2, we study the linear stability of the disordered phase for an active polar fluid (see Supplementary Notes 1). The velocity and polarity fields are expressed in Fourier bases with wavevectors $(q_x, q_y)$. For each Fourier component, we identified two relevant modes that evolve in time, up to linear order in perturbations, according to the dynamical equations expressing force balance and incompressibility (see Supplementary Notes 2, eq (18)). These equations can be explicitly integrated in time, resulting in an exponential evolution for the two relevant modes with an initial amplitude set by initial conditions. We choose each of these amplitudes to obey a normal distribution with zero mean and standard deviation set to $q = \sqrt{q_x^2 + q_y^2}$. For each Fourier component, these two modes are related to the Fourier components of the velocity field and the polarity field through Eqs. (16) and (17), respectively. This allows the velocity and polarity field, such as that shown in Figure 4C, to be computed at every time point.

To produce Figure 4C, we set the units $\chi = \gamma = k = 1$. Then, we chose the rest of parameters: $\lambda_\parallel = \frac{1}{2}$, $\lambda_\perp = 2$, $k = 0.1$, $T_0 = 20$, and $\eta = 10$. The time was set to 0.5, the integration domain was a square with a lateral size L = 800 and the Fourier series were truncated at 60 modes in each direction. In Figure 4C, we show the velocity field in a square subdomain of size L/2.

**Particle-based simulation of active polar laning:**

We used a particle-based model of active cells to simulate the laning transition resulting from anisotropic substrate friction. Cells were represented as two-dimensional self-propelled discs of radius $R$, with position $\mathbf{x}$, polarity $\mathbf{p}$, and normalized velocity $\hat{\mathbf{v}} = \mathbf{v}/v_o$, with $v_0$ an arbitrary reference velocity. Each cell *i* is governed by the overdamped equation of motion

$$\widehat{\Lambda}_s \hat{\mathbf{v}}_i = \mathbf{p}_i - \nabla_i \widehat{U}(\mathbf{x}_i)$$

with anisotropic viscous friction $\widehat{\Lambda}_s = \hat{\mathbf{x}}\hat{\mathbf{x}}^T + h_s(\mathbf{I} - \hat{\mathbf{x}}\hat{\mathbf{x}}^T)$ and a soft interaction potential, $\widehat{U}(\mathbf{x}_i) = \sum_j \max(0, \frac{k_c}{2R}(2R - \|\mathbf{x}_i - \mathbf{x}_j\|)^2)$, with stiffness $k_c$. Cell polarity evolves as a non-linear stochastic process with velocity reinforcement $k_s$ and diffusivity $D_p$,

$$d\mathbf{p}_i = [k_s \hat{\mathbf{v}}_i - (k_s + 1)\|\hat{\mathbf{v}}_i\|^2 \mathbf{p}_i] D_p\, dt + \sqrt{2 D_p}\, d\mathbf{W}_i$$

where $\mathbf{W}_i$ is a two-dimensional Wiener process. Equations of motion for translation and polarization were integrated using, respectively, an explicit Euler scheme and the Euler-Maruyama method, with a time step $\Delta t = 0.015\, R/v_0$. Simulations ran until $t = 2000\, R/v_0$, when macroscopic coarsening had fully developed, and the reported quantities remained constant or changed only very slowly. Quasi-steady state measures reported in Figure 4D-I were extracted over the final time window of $t' = 200\, R/v_0$. Simulations were performed in a rectangular domain of size $x = 200\, R$ and $y = 1000\, R$, with periodic boundary conditions. To initialize simulations, $N$ particles were placed in a hexagonal grid with overlapping potentials at a fixed number density $n$ (3820 cells/mm$^2$), corresponding to a relative density of $\rho = \pi R^2 n = 1.2$, or $N = 75{,}696$. To prevent grid artefacts, we introduced a small variability in the cell radius, such that the cell radius is drawn from a normal distribution with mean $R$ and standard deviation $\sigma_R = 0.05\, R$. Cell polarity was initialized at $\mathbf{0}$. To obtain robust statistics, we performed multiple runs for each configuration, where each run was initialized with a different pseudo-random number seed. The number of repeats was 3, unless stated otherwise in the Figure caption. All model parameters are reported in Supplementary Table 1.

The guidance order parameter was calculated as $S = \langle \cos(2\theta_i) \rangle_{i,t}$, where the angle of the polarization $\theta_i$ was averaged over all cells $i \in [1, N]$ and over the final time window $t'$. Consistent with the experimental analysis of microscopy images, we binned the output of these simulations in a rectangular grid with window size $L_b = 2.6$ cell radii, in which each window value was o by averaging velocity and polarity from 4 grid windows of size $2L_b$, shifted with displacements $(\pm L_b, \pm L_b)$. Based on the binned value of $v_x$, we computed the velocity correlation function in the same way as in the experiments. From this function, the x correlation length $L_x$ was obtained as $\frac{1}{L_x} = -\lim_{\Delta x \to 0} \frac{d\, C(\Delta x)}{dx}$, where $C(\Delta x)$ is the average of the x correlation function over $t'$. A detailed description and analysis of the particle-based model can be found in Supplementary Notes 5.

## Data availability

The data that support the findings of this study and that are not in the paper and Supplementary Material file, have been deposited in Zenodo at https://doi.org/XXXXX

## References

48. Duclos, G. *et al.* Controlling Confinement and Topology to Study Collective Cell Behaviors. in *Cell Migration: Methods and Protocols, Methods in Molecular Biology* (ed. Gautreau, A.) **1749**, 387–399 (Humana Press, 2018).

49. Poujade, M. *et al.* Collective migration of an epithelial monolayer in response to a model wound. *Proc. Natl. Acad. Sci. U. S. A.* **104**, 15988–93 (2007).

50. Schindelin, J. *et al.* Fiji: an open-source platform for biological-image analysis. *Nat. Methods* **9**, 676–682 (2012).

51. Rasband, W. S. ImageJ v1.46b. *ImageJ v1.46b (US Natl. Inst. Heal. Bethesda, Maryland, 1997-2012)*

52. Preibisch, S., Saalfeld, S. & Tomancak, P. Globally optimal stitching of tiled 3D microscopic image acquisitions. *Bioinformatics* **25**, 1463–1465 (2009).

53. Petitjean, L. *et al.* Velocity Fields in a Collectively Migrating Epithelium. *Biophys. J.* **98**, 1790–1800 (2010).

54. Sveen, J. K. An introduction to MatPIV v. 1.6.1. *Eprint Ser. Dept. Math. Univ. Oslo, "Mechanics Appl. Math.* **ISSN 0809**-, (2004).